\documentclass[reprint,amsmath,amssymb,aps,superscriptaddress,nofootinbib,longbibliography]{revtex4-1}

\usepackage{graphicx}
\usepackage{natmove}
\usepackage{hyperref}
\hypersetup{colorlinks,allcolors=blue}
\usepackage[capitalize]{cleveref}
\usepackage{newtxtext}
\usepackage[varvw]{newtxmath}

\input{macros.sty}

\begin{document}

\title{Feasibility of lasing in the GaAs Reststrahlen band
with HgTe multiple quantum well laser diodes}

\author{Aleksandr Afonenko}
\email{afonenko@bsu.by}
\affiliation{Belarusian State University, 220030 Minsk, Belarus}

\author{Dmitry Ushakov}
\affiliation{Belarusian State University, 220030 Minsk, Belarus}

\author{Georgy Alymov}
\affiliation{Laboratory of 2d Materials for Optoelectronics, Moscow Institute of Physics and Technology, Dolgoprudny 141700, Russia}

\author{Aleksandr Dubinov}
\email{sanya@ipmras.ru}
\affiliation{Institute for Physics of Microstructures, Russian Academy of Sciences, 603950 Nizhny Novgorod, Russia}

\author{Sergey Morozov}
\affiliation{Institute for Physics of Microstructures, Russian Academy of Sciences, 603950 Nizhny Novgorod, Russia}

\author{Vladimir Gavrilenko}
\affiliation{Institute for Physics of Microstructures, Russian Academy of Sciences, 603950 Nizhny Novgorod, Russia}

\author{Dmitry Svintsov}
\affiliation{Laboratory of 2d Materials for Optoelectronics, Moscow Institute of Physics and Technology, Dolgoprudny 141700, Russia}
\email{svintcov.da@mipt.ru}
             
\date{}             
             
\begin{abstract}
Operation of semiconductor lasers in the 20--50 $\mu$m wavelength range is hindered by strong non-radiative recombination in the interband laser diodes, and strong lattice absorption in GaAs-based quantum cascade structures. Here, we propose an electrically pumped laser diode based on multiple HgTe quantum wells with band structure engineered for Auger recombination suppression. Using a comprehensive model accounting for carrier drift and diffusion, electron and hole capture in quantum wells, Auger recombination, and heating effects, we show the feasibility of lasing at  $\lambda = 26...30$ $\mu$m at temperatures up to 90 K. The output power in the pulse can reach up to 8 mW for microsecond-duration pulses.
\end{abstract}

\maketitle
\section{\label{sec:introduction}Introduction}

The development of compact sources of far-infrared (FIR) radiation is among the most important problems in modern semiconductor physics. Such sources are demanded by many applications, including gas and solid-state spectroscopy and environment monitoring~\cite{2015/Neumaier/Analyst/Terahertz,2015/Hochrein/JIMTHzW/Markets,2017/Dhillon/JPhysD/The2017}. Unipolar quantum cascade lasers (QCLs) based on GaAs/AlGaAs or InGaAs/InAlAs/InP are the most popular compact FIR sources in the 50--300 $\mu$m region and below 20 $\mu$m. However, the 20--50 $\mu$m range is hardly accessible by existing QCLs due to the strong optical absorption induced by polar optical phonons~\cite{2015/Vitiello/OptExpr/Quantum} (reststrahlen band). There exist only a few works on QCLs operating within the 20--50 $\mu$m range~\cite{2014/Ohtani/APL/Double,2019/Loghmari/ElectrLet/InAs-based,2016/Ohtani/ACSPhotonics/Far-Infrared}.

The problem of lattice absorption is absent in \AIIBVI{} materials, in particular, HgCdTe, where the optical phonon frequencies correspond to $\sim 70$ $\mu$m wavelength. Studies on bulk HgCdTe solid solutions have been conducted for more than 50 years, and extensive knowledge has been accumulated about the properties and technology of these compounds where the bandgap varies between zero and 1.6 eV depending on the composition. This material is widely used for mid-infrared photodetectors and detector arrays (see \Cite{2005/Rogalski/Reports/HgCdTe} and references therein). The interest to HgCdTe material system re-emerged in the latest decade after prediction~\cite{Bernevig} and observation~\cite{Mollenkamp} of various topological electronic phases, both in bulk material and its quantum wells. Depending on Cd fraction, quantum well thickness, magnetic and electric field, temperature, and hydrostatic pressure, CdHgTe can be either a topological insulator, a Dirac semimetal, or a conventional narrow-gap semiconductor~\cite{2016/Teppe/NatComm/Temperature,Kadykov_PRL}.

Among a variety of electronic phases realised in CdHgTe quantum wells, the most attractive one for far-infrared lasing is the narrow-gap semiconductor with symmetric quasi-relativistic electron-hole dispersion~\cite{2017/Ruffenach/APLMat/HgCdTe}. Realisation of such dispersion enforces strong suppression of non-radiative Auger recombination~\cite{lead-salt,Emtage_1976,Alymov_PRB} that poses the main obstacle for maintaining interband population inversion in narrow-gap semiconductors. Generally, the rate of Auger recombination displays an activation-type dependence on temperature, $R_A \propto \exp(-E_{\rm th}/kT)$. The threshold energy $E_{\rm th}$ is proportional to the band gap, $E_{\rm th} = \gamma E_g$, with $\gamma$ being related to the band dispersion~\cite{Abakumov-nonradiative}. Very recently, we have shown that $\gamma$ can exceed unity in HgTe quantum wells with thickness $d\sim 6$ nm~\cite{2020/Alymov/ACSPhot/Fundamental}, while in conventional semiconductors with parabolic bands $\gamma$ is always below $1/2$.

There is continued experimental evidence for suppression of non-radiative Auger recombination in narrow quantum wells of pure HgTe~\cite{2016/Morozov/APL/Long,rumyantsev2018effect}. In particular, stimulated emission up to $\lambda = 20$ $\mu$m was recently achieved at cryogenic temperatures~\cite{2017/Morozov/APL/Stimulated}, and the maximum temperature of stimulated emission at $\lambda = 3.7$ $\mu$m was raised to 240 K~\cite{Kudryavtsev_TempLimitations}. It is remarkable that such strong Auger suppression was overlooked for about half-a-century study of HdCdTe for lasing applications\cite{1966/Melngailis/APL/SPONTANEOUS,1993/Arias/SemScienTechn/HgCdTe}. Most of CdHgTe-based quantum well structures studied previously were either wide~\cite{1999/Bleuse/JCresGr/Laser}, or contained large fraction of cadmium~\cite{1998/Vurgaftman/OE/Hight}, which made Auger recombination relatively strong.

Motivated by the confirmed feasibility of stimulated emission and suppressed Auger recombination in optically pumped HgTe quantum wells, here we study the feasibility of electrically pumped lasing. The proposed structure is based on multiple HgTe quantum wells of $5.2$ nm thickness placed in Cd$_x$Hg$_{1-x}$Te barrier layer. The quantum well band gap is tuned for lasing at $\lambda = 26...30$ $\mu$m. By employing a comprehensive numerical model that takes into account drift and diffusion of carriers in barrier layers, capture into quantum wells, non-radiative and radiative recombination, and heating of active region, we show the feasibility of lasing at temperatures up to $\sim 90$ K.  We show that the main limiting factors for temperature performance are the residual Auger recombination in the quantum wells and Drude absorption in highly doped injecting regions. While \emph{electrically} pumped FIR HgCdTe lasers are yet to be experimentally realized, there have been theoretical proposals of a HgCdTe QCL~\cite{2020/Ushakov/OE/HgCdTe} and a FIR HgCdTe laser based on difference frequency generation~\cite{HgCdTe_DFG}.
\begin{widefig}{structure}{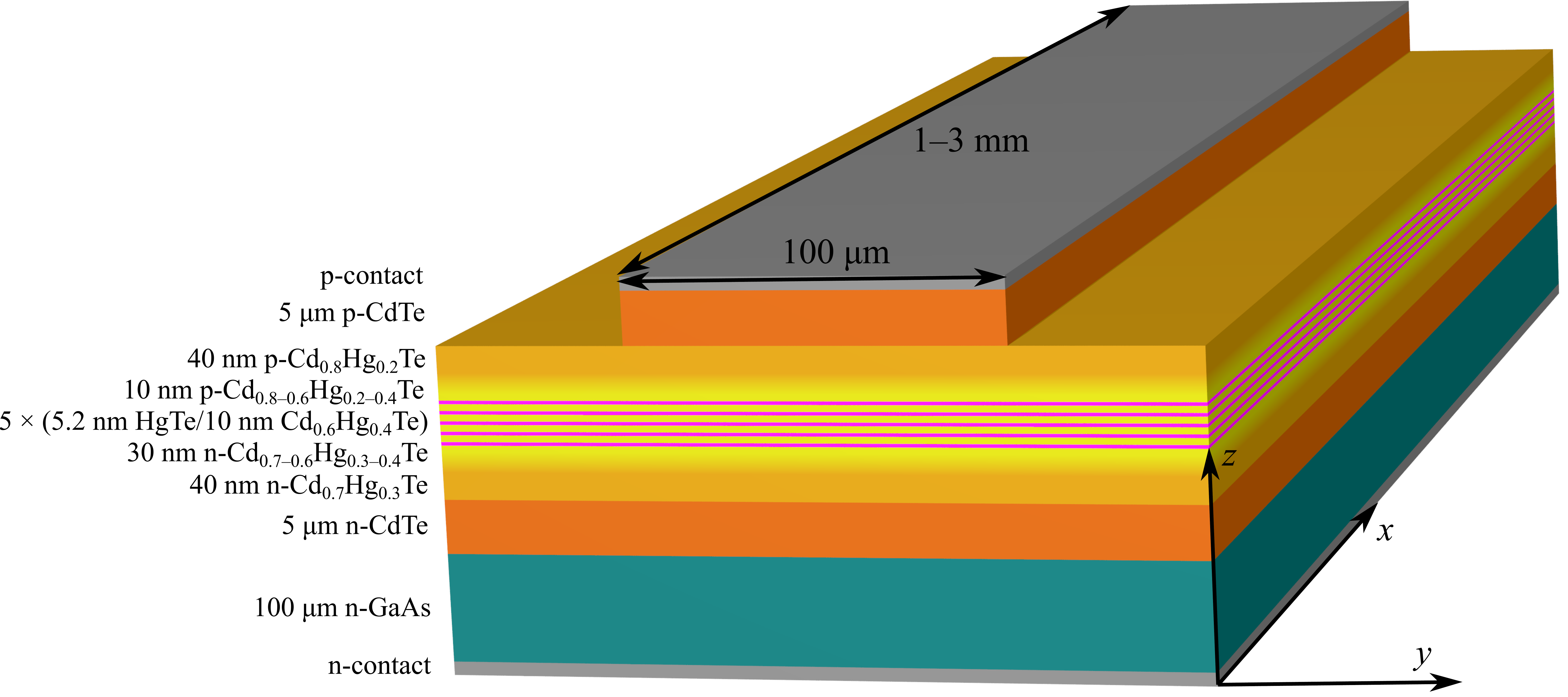}
Schematic of the proposed laser structure, consisting of the active region (HgTe quantum wells, magenta + Cd$_{0.6}$Hg$_{0.4}$Te barriers, yellow), waveguide layers (light orange), cladding layers (orange), substrate (green), and metal contacts (gray).
\end{widefig}

\begin{table*}[!t]
\caption{\label{tab:parameters}
Parameters of heterostructure layers: layer number (\#), thickness ($d$), composition, doping, electron ($\mu_n$) and hole ($\mu_p$) mobilities, bandgap ($E_g$), CCCH ($C_{nnp}$) and CHHH ($C_{npp}$) Auger coefficients, electron ($\sigma_n$) and hole ($\sigma_p$) absorption cross sections, and real part of the refractive index ($n'$). All the quantities are for 3D materials. In the graded-composition waveguide layers, the parameters vary linearly between their values in adjacent layers. 
}

\begin{ruledtabular}
\begin{tabular}{*{12}c}

\# & $d$  & Composition & Doping  & $\mu_n$  & $\mu_p$ & $E_g$ & $C_{nnp}$  & $C_{npp}$ & $\sigma_{n}$ & $\sigma_{p}$  & $n'$\cr
 &  (nm) &  &  (cm$^{-3}$) & (cm$^2$/V$\cdot$s) & (cm$^2$/V$\cdot$s) &  (eV) &  (cm$^6$/s) & (cm$^6$/s) &  (cm$^2$) &  (cm$^2$) & \cr
\multicolumn{12}{c}{Substrate}\\
1 &  & GaAs & $n: 8\cdot 10^{17}$ & 10800 & 850 & 1.51 & $1.9\cdot 10^{-31}$ & $1.2\cdot 10^{-30}$ & $4.2\cdot 10^{-16}$ & $9.1\cdot 10^{-17}$ & 0 \cr \hline
\multicolumn{12}{c}{$n$-cladding layer}\\
2 & 5000 & CdTe & $n: 10^{16}$ & 1150 & 11 & 1.60 & $9.0\cdot 10^{-43}$ & $4.4\cdot 10^{-30}$ & $2.2\cdot 10^{-15}$ & $6.9\cdot 10^{-15}$ & 2.55\cr \hline
\multicolumn{12}{c}{$n$-waveguide layers}\\
3 & 40 & Cd$_{0.7}$Hg$_{0.3}$Te & $n: 10^{16}$ & 2100 & 21 & 1.00 & $5.5\cdot 10^{-33}$ & $6.9\cdot 10^{-28}$ & $2.2\cdot 10^{-15}$ & $3.5\cdot 10^{-15}$ & 2.81\cr
4 & 30 & graded \cr 
5 & 10 & Cd$_{0.6}$Hg$_{0.4}$Te & $n:10^{16}$ & 2550 & 25 & 0.81 & $2.1\cdot 10^{-30}$ & $4.0\cdot 10^{-27}$ & $2.5\cdot 10^{-15}$ & $2.8\cdot 10^{-15}$ & 2.90\cr\hline
\multicolumn{12}{c}{active region}\\
6 & 5.2 & QW: HgTe & 0 & 28 000 & 80 & -0.26 & $9.0\cdot 10^{-25}$ & 0 & $2.3\cdot 10^{-16}$ & $2.5\cdot 10^{-16}$ & 3.75\cr
7 & 10 & Cd$_{0.6}$Hg$_{0.4}$Te & 0 & 28 000 & 280 & 0.81 & $2.1\cdot 10^{-30}$ & $4.0\cdot 10^{-27}$ & $2.3\cdot 10^{-16}$ & $2.5\cdot 10^{-16}$ & 2.91\cr
\multicolumn{12}{c}{ + 3 + 1/2 periods (2 layers in period)}\\
\hline
\multicolumn{12}{c}{$p$-waveguide layers}\\
15 & 10 & Cd$_{0.6}$Hg$_{0.4}$Te & $p:10^{16}$ & 2550 & 25 & 0.81 & $2.1\cdot 10^{-30}$ & $4.0\cdot 10^{-27}$ & $2.5\cdot 10^{-15}$ & $2.8\cdot 10^{-15}$ & 2.91\cr
16 & 10 & graded\cr
 17 & 40 & Cd$_{0.8}$Hg$_{0.2}$Te & $p: 10^{16}$ & 1700 & 17 & 1.19 & $6.5\cdot 10^{-36}$ & $1.3\cdot 10^{-28}$ & $2.1\cdot 10^{-15}$ & $4.4\cdot 10^{-15}$ & 2.73\cr
\hline
\multicolumn{12}{c}{$p$-cladding layer}\\
18 & 5000 & CdTe & $p: 10^{15}$ & 6300 & 60 & 1.60 & $9.0\cdot 10^{-43}$ & $4.4\cdot 10^{-30}$ & $3.9\cdot 10^{-16}$ & $1.3\cdot 10^{-15}$ & 2.56\cr
\end{tabular}
\end{ruledtabular}

\end{table*}

\section{\label{sec:methods}Methods}
We consider a multiple quantum well laser structure grown on a (013)-oriented $n$-GaAs substrate and consisting of an active region (five 5.2 nm HgTe QWs interleaved with 10 nm Cd$_{0.6}$Hg$_{0.4}$Te barriers), 70 nm Cd$_x$Hg$_{1-x}$Te waveguide layers ($x$ varies between 0.6 and 0.7), and 5 $\mu$m CdTe cladding layers (Fig.~\ref{fig:structure}, Table~\ref{tab:parameters}).

\begin{widefig}{bands_and_concentrations}{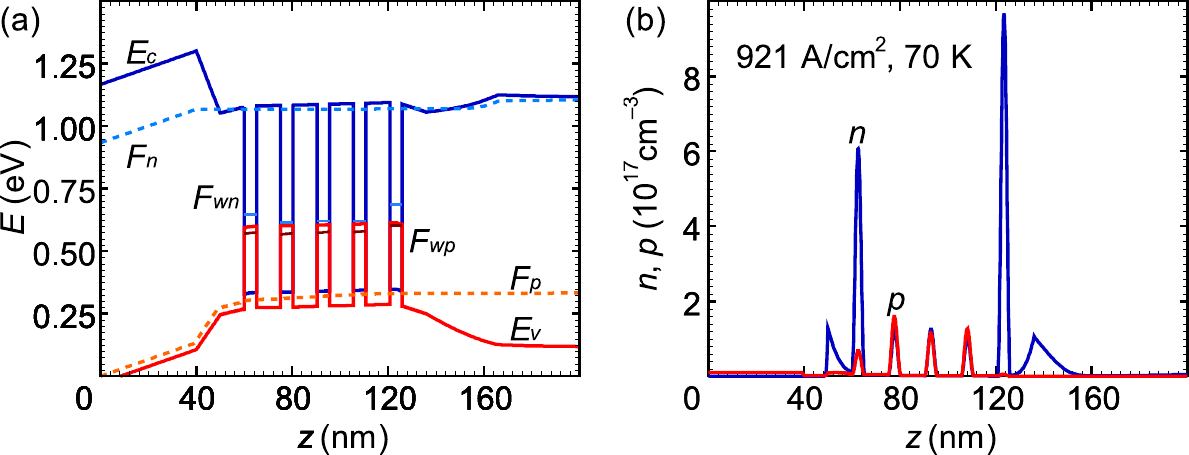}
(a) Band diagram and (b) distribution of carrier densities inside the active region of the simulated heterostructure, calculated at 70 K, 1.1 V bias, 921 A/cm$^2$ drive current density (above the lasing threshold), and $S^{(2D)}=4.3 \times 10^{10}$ cm$^{-2}$  photon density. $F_{wn}$, $F_{wp}$ ($F_n$, $F_p$) are the quasi-Fermi levels of localized (delocalized) carriers. 
\end{widefig}
\begin{widefig}{optical_properties}{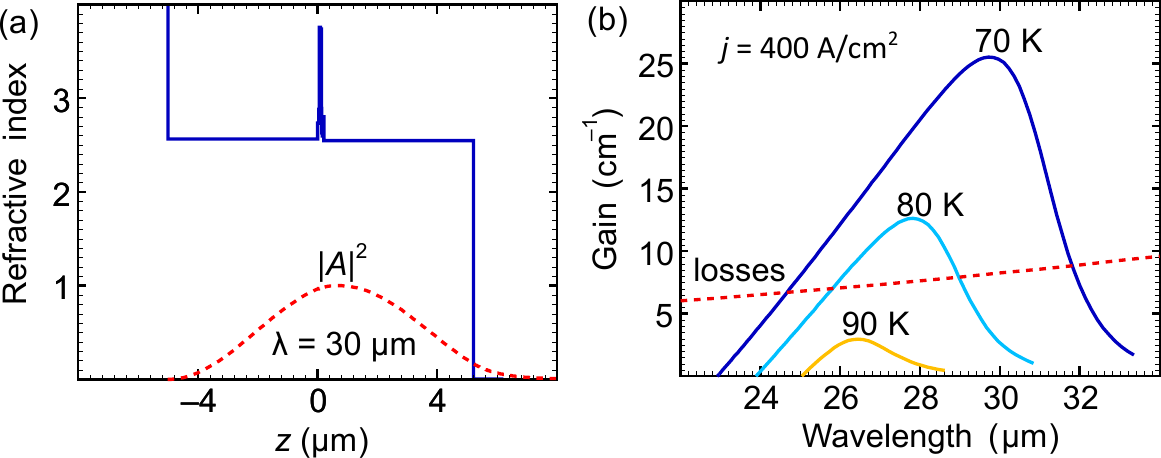}
(a) Spatial distribution of the refractive index across the simulated heterostructure at $\lambda = 30$ $\mu$m and the squared electric field of the ground TE mode. The total optical confinement factor of 5 quantum wells is $\Gamma = 0.0043$, the effective refractive index is $n_{\rm eff}$ = 2.24. 
(b) Gain spectra (unsaturated: photon density $S^{(2D)}=0$) for the TE$_0$ mode at temperatures $T = 70$, 80, 90 K. The corresponding drive current density is $j = 400$ A/cm$^2$. The estimated total losses are also shown with dashed line.
\end{widefig}

This structure was simulated using a distributed drift-diffusion model based on one-dimensional (1D) Poisson's equation and continuity equations for electrons and holes with taking into account the  carrier capture and  escape processes~\cite{2014/Afonenko/FTP/Current}. Both radiative and Auger recombination were included in the model.

The bandstructure parameters were obtained from the eight-band $\vec{k} \cdot \vec{p}$ method~\cite{2005/Novik/PhysRevB/Band,2016/ALESHKIN/PhysB/Effect}. Quantum well depths for electrons and holes were calculated according to Refs.~\citenum{2005/Novik/PhysRevB/Band,1989/Van_de_Walle/PhysRevB/Band}. Carrier mobilities were interpolated from the experimental data of \Cite{1972/Scott/JAP/Electron}. The calculated parameters of each heterostructure layer are collected in Table~\ref{tab:parameters}.

Internal optical losses and refractive index were calculated within the multioscillator Lorentz-Drude model including both phonon and free-carrier contributions~\cite{1983/Mroczkowski/JAP/Optical,1990/Laurenti/JAP/Temperature,1974/Grynberg/PhysRevB/Dielectric}.

Distribution of the photon density across the resonator was found from the Bouguer--Lambert--Beer law. 

Heating effects were taken into account by solving the 1D heat equation in the direction perpendicular to the heterostructure layers (we consider pulsed operation, when in-layer heat transfer during a pulse is negligible). We also included the temperature dependence of the bandgap in our model.

A more detailed account of our methods we used is presented in Appendices \ref{sec:drift-diffusion}, \ref{sec:resonator-heating}, and \ref{sec:Auger}.
\section{\label{sec:results}Results and discussion}
\begin{widefig}{time_evolution}{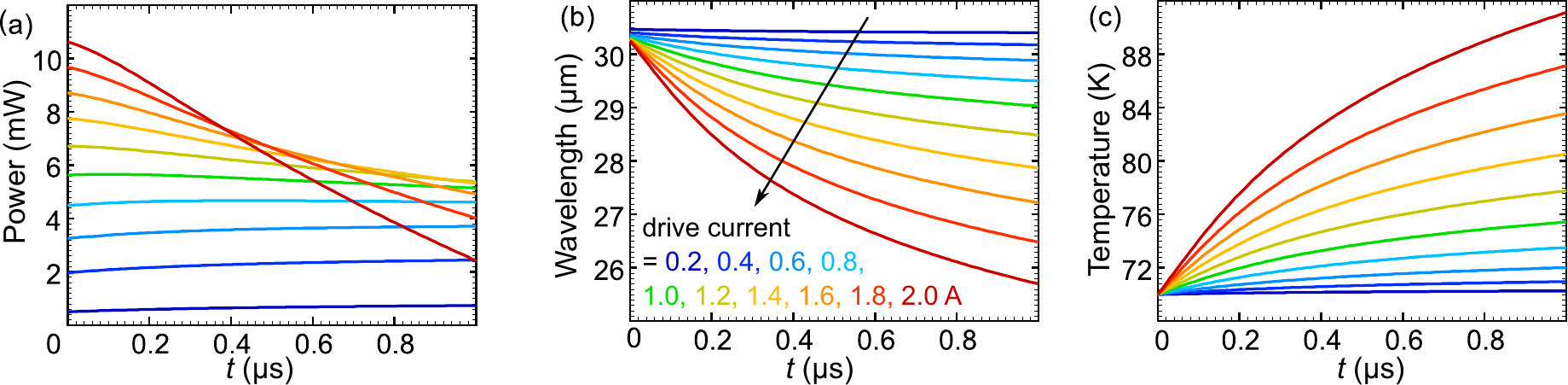}
Time evolution of (a) output power, (b) lasing wavelength, and (c) temperature of the active region during a 1 $\mu$s pulse at different drive currents. The heat sink temperature is 70 K.
\end{widefig}
\begin{widefig}{spatial_distribution}{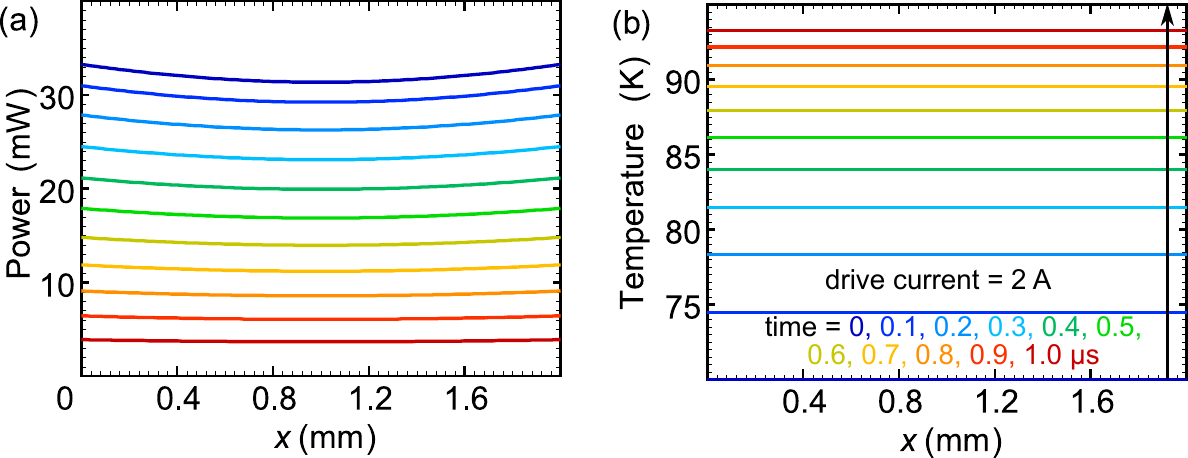}
Spatial distribution of (a) output power and (b) temperature of the active region across a 2-mm-long resonator at different times since the beginning of the pulse. The drive current is 2 A, the heat sink temperature is 70 K.
\end{widefig}

An example of calculated band diagrams and carrier distributions are shown in Fig.~\ref{fig:bands_and_concentrations}. The quasi-Fermi level difference inside the QWs $F_{wn} - F_{wp}$ is much smaller than in the barriers, meaning most of the carrier energy is dissipated into heat upon carrier capture into the QWs. One could try to increase the laser efficiency by lowering the barrier height; however, this is undesirable because it would boost Auger recombination in the barriers~\cite{2019/Aleshkin/JPhysCond/Threshold}.

Another important observation is that the population inversion is distributed nonuniformly across the active region. 
To obtain larger gain, it is advantageous to use several quantum wells instead of a single one.
Due to the low mobility of holes, the level of excitation of quantum wells decreases with distance from the $p$-emitter.
According to our calculations, $N_{\rm QW} = 5$ is the optimal number of QWs, beyond which no significant increase in modal gain can be achieved.

Nonuniform carrier distribution can also appear in waveguide layers. We found that in constant-composition waveguides, small density and mobility of holes lead to carrier accumulation near the $p$-cladding, with carrier density exceeding the nominal doping level by more than an order of magnitude. Putting a thin graded-composition Cd$_x$Hg$_{1-x}$Te layer near the $p$-cladding with $x$ varying linearly between 0.8  and 0.6 eliminates carrier accumulation (Fig.~\ref{fig:bands_and_concentrations}b), which explains why experimentally measured lasing thresholds in structures with graded-composition layers were lower than in those with constant-composition layers~\cite{1999/Bleuse/JCresGr/Laser}.

Bandgap difference between the adjacent waveguide layers (Cd$_x$Hg$_{1-x}$Te with $0.7 \leq x \leq 0.8$  and the barriers ($x = 0.6$) results in good current confinement and negligible carrier leakage into the waveguide (electron quasi-Fermi level $F_n$ in the $p$-doped waveguide layer and hole quasi-Fermi level $F_p$ in the $n$-doped waveguide layer lie deep within the bandgap, Fig.~\ref{fig:bands_and_concentrations}~a).

Figure~\ref{fig:optical_properties}a shows the good optical confinement of the TE$_0$ mode provided by the heavily doped GaAs substrate (with nearly zero real part of the refractive index, see Fig.~\ref{fig:optical_properties}a) on one side of the heterostructure and a metal contact on the other side. The calculated total optical confinement factor of 5 QWs is $\Gamma = 0.0043$. The reflection coefficients of the facets for the TE$_0$ mode were estimated using eigenmode decomposition~\cite{1972/Ikegami/IEEEJournQE/Reflectivity} as $r_1$, $r_2 = 0.5$ at 30 $\mu$m, which is much higher than the Fresnel reflection coefficient for a plane wave (0.2).

\begin{widefig}{light-current}{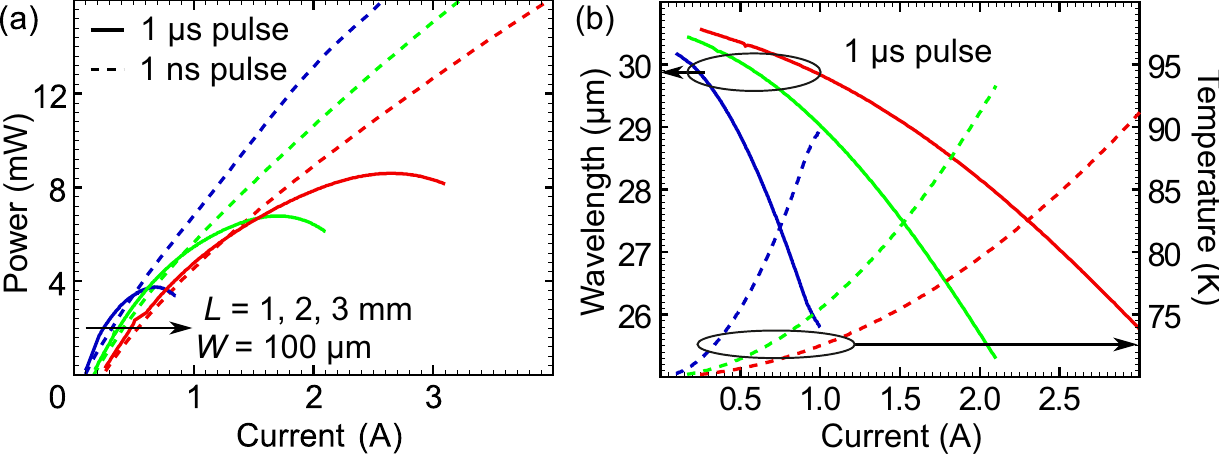}
Calculated (a) Output power (b)  lasing wavelength and temperature of the active region at the end of 
1 $\mu$s pump pulse vs drive current for different resonator lengths $L$. Resonator width is 100 $\mu$m, reflection coefficients of the facets are $r_1 = r_2 = 0.5$. The heat sink temperature is 70 K.
\end{widefig}

The good current and optical confinement result in modal gain exceeding the total optical loss (estimated as 7--12 cm$^{-1}$ in the simulated structure) up to 
$\sim 90$ K (Fig.~\ref{fig:optical_properties}b), thus allowing laser operation slightly above the boiling point of liquid nitrogen. The gain maximum is shifted to shorter wavelengths with increasing temperature due to the temperature dependence of the bandgap.

When discussing the operating temperature, one should take into account that the temperature of the active region can exceed the heat sink temperature, especially at large drive currents. In FIR lasers, heating issues are aggravated by the tight restrictions on the cladding doping level to avoid prohibitive optical losses. In our simulations, we used a doping level of $10^{15}$ cm$^{-3}$ for the $p$-cladding, resulting in resistance as high as 15 $\Omega$ for a 2 mm-long and 100 $\mu$m-wide resonator. Fig.~\ref{fig:time_evolution} shows that, for a drive current of 2.0 A, the temperature of active region grows by 23 K during a 1 $\mu$s pulse, the output power reduces from 10.6 to 2.4 mW, and the lasing wavelength shifts from 30.2 to 25.7 $\mu$m.

On the other hand, good spatial uniformity of lasing and heating is maintained throughout the pulse. Fig.~\ref{fig:spatial_distribution} shows that the lasing power and temperature are almost uniform across the resonator for reflection coefficients of the facets $r_1 = r_2 = 0.5$. However, if anti-reflective and highly reflective coatings are used, spatial nonuniformity may become pronounced, with the lasing power at the front facet being considerably higher than at the back facet.

Finally, we calculated the output power of the laser under study. Figure~\ref{fig:light-current} shows the output power, the temperature of the active region, and the lasing wavelength vs the drive current at three values of the resonator length: $L = 1$, 2, 3 mm. The resonator width is $W = 100$ $\mu$m.
Maximum output power increases with the cavity length which results from weaker heating in longer cavity devices. At 1 $\mu$s pulse length (quasi-CW mode), the maximum output power is 8.6 mW  for $L = 3$ mm.
During the action of the pump pulse, the active region is heated and the radiation wavelength changes. As can be seen from the figure, the pulse-induced change in the  wavelength and temperature are $\sim$ 4 $\mu$m and $\sim$ 20 K, respectively (see Fig.~\ref{fig:light-current}b).
Heating issues can be mitigated with shorter pulses, which allow to achieve several tens of mW output power (see Fig.~\ref{fig:light-current}a). It should be noted that a higher power in the initial part of the watt-ampere characteristics is realised at longer pump pulses. This is due to the heating of the active region during the pulse and a decrease in the laser wavelength, which leads to a decrease in internal losses (see Fig.~\ref{fig:time_evolution}).
\section{\label{sec:conclusions}Conclusions}

We have theoretically demonstrated the feasibility of lasing under electrical pumping of HgTe quantum wells in the $26...30$ $\mu$m wavelength range. This range of wavelengths lies in the phonon absorption (reststrahlen) band of GaAs-based compounds and is thus unattainable for existing quantum cascade lasers. The model used takes into account all necessary aspects of laser action, including drift and diffusion of carriers in the barrier layers, electron and hole capture of quantum wells, radiative and non-radiative Auger recombination, as well as heating of active region. According to the results of simulations, the laser action is possible at temperatures up to 90 K. At 70 K heat sink temperature, the average output power during a 1 $\mu$s pulse can reach $\sim$ 9 mW. Feasibility of lasing at such long wavelengths and liquid nitrogen temperature is due to suppression of Auger recombination in narrow HgTe quantum wells which, in turn, is inherited from symmetric quasi-relativistic electron-hole dispersion.
\begin{acknowledgments}
The work of DS and GA was supported by Russian Foundation for Basic Research, grant \# 19-37-70031. AD, SM, and VG acknowledge the support of Russian Science Foundation, grant \# 17-12-01360.
\end{acknowledgments}
\appendix

\section{\label{sec:drift-diffusion}Distributed drift-diffusion model}
To calculate the band diagrams, we used a model based on 1D Poisson's equation and continuity equations for electrons and holes with taking into account nonuniform distribution of population inversion across the heterostructure~\cite{2014/Afonenko/FTP/Current}. The main quantities in our distributed model are the electric potential $\varphi$ and the electron and hole quasi-Fermi levels $F_n$, $F_p$, all of which depend on the coordinate $z$ along the growth direction. These quantities are found by numerically integrating Poisson's equation together with the continuity equations for the electron and hole current densities $j_n$, $j_p$:
\begin{eq}{Poisson-continuity}
\frac{d^2 \varphi}{dz^2} &= -\frac{e}{\epsilon \epsilon_0}(p - n + N_d - N_a),\\
\frac{\partial n}{\partial t} - \frac{1}{e}\frac{\partial j_n}{\partial z} &= \frac{\partial p}{\partial t} + \frac{1}{e}\frac{\partial j_p}{\partial z} = - R - v_g G S^{(3D)},\\
\quad j_n &= \mu_n n_b \frac{\partial F_n}{\partial z}, \quad j_p = \mu_p p_b \frac{\partial F_p}{\partial z}.
\end{eq}
Here, $e$ is the elementary charge, $n$ ($p$) is the electron (hole) density, $N_a$ ($N_d$) is the density of ionized acceptors (donors), $\epsilon$ is the relative permittivity of CdHgTe, $\epsilon_0$ is the vacuum permittivity, $R$ is the interband (nonradiative and spontaneous) recombination rate, $G$ is the modal gain, $S^{(3D)}$ is the 3D photon density, $v_g$ is the group velocity of the lasing mode, and $\mu_n$ ($\mu_p$) is the electron (hole) mobility.

The total electron and hole densities $n$, $p$ consist of the densities of delocalized carriers, contributing to the electric current:
\begin{eq}{delocalized_densities}
n_b &= N_c^{(3D)} \Phi_{1/2}\left(\frac{F_n - E_c}{kT}\right),\\
p_b &= N_v^{(3D)} \Phi_{1/2}\left(\frac{E_v - F_p}{kT}\right),
\end{eq}
and densities of carriers localized in the quantum wells:
\begin{eq}{localized_densities}
n_w &= N_c^{(2D)} \sum_i \ln \left[ 1 + \exp\left(\frac{F_{wn} - E_{ci}}{kT}\right) \right] \abs{\psi_i(z)}^2,\\
p_w &= \sum_i N_{vi}^{(2D)} \ln \left[ 1 + \exp\left(\frac{E_{vi} - F_{wp}}{kT}\right) \right] \abs{\psi_i(z)}^2.
\end{eq}
Here, $N_c^{(3D)}$, $N_v^{(3D)}$ ($N_c^{(2D)}$, $N_{vi}^{(2D)}$) are the 3D (2D) effective densities of states for electrons and holes, $k$ is the Boltzmann constant, and $T$ is the temperature. Logarithms and complete Fermi-Dirac integrals of order $1/2$, $\Phi_{1/2}$, stem from integrating Fermi-Dirac distributions with quasi-Fermi levels $F_n$, $F_p$ (for delocalized carriers) and $F_{wn}, F_{wp}$ (for localized carriers). Carrier states at $i$th subband in a quantum well are described by wavefunctions $\psi_i(z) = \sqrt{2/d}\sin\left(i\pi z/d \right)$, assuming the well spans the region $0 < z < d$. Subband edges $E_{ci}$, $E_{vi}$ and effective densities of states were found from the eight-band $\vec{k} \cdot \vec{p}$ method~\cite{2005/Novik/PhysRevB/Band,2016/ALESHKIN/PhysB/Effect}.

3D photon density inside the $m$-th well is connected to the 2D photon density $S^{(2D)}$ by the optical confinement factor $\Gamma_m$:
\begin{eq}{photon_density}
S_m^{(3D)} = \frac{\Gamma_m}{d} S^{(2D)}.
\end{eq}
The dynamics of the localised carriers and 2D photon density was described by standard rate equations \cite{2014/Afonenko/FTP/Current}:
\begin{eq}{photon_rate_eqs}
\frac{d n_{wm}^{2D}}{dt} &= \frac{n_{bm}^{2D}}{\tau_{{\rm cap}, n}} \left(1 - e^{-\Delta F_{nm}/kT} \right) - R_{m}^{(2D)} - v_g G_m S^{(2D)},\\
\frac{d p_{wm}^{2D}}{dt} &= \frac{p_{bm}^{2D}}{\tau_{{\rm cap},p}} \left(1 - e^{-\Delta F_{pm}/kT} \right) - R_{m}^{(2D)} - v_g G_m S^{(2D)},\\
\frac{d S^{(2D)}}{dt} &= v_g \left(\sum_m G_m - \alpha_{\rm tot} \right) S^{(2D)} + \beta \sum_m R_{{\rm sp}\,m}^{(2D)},
\end{eq}
where $n_{wm}^{2D}$ ($p_{wm}^{2D}$) is the 2D density of electrons (holes) localized in the $m$th well, $n_{bm}^{2D}$ ($p_{bm}^{2D}$) is the density of delocalized carriers in the $m$th well, $\Delta F_{nm}$ ($\Delta F_{pm}$) is the quasi-Fermi level difference between delocalized and localized electrons (holes), $\tau_{{\rm cap}, n}$ ($\tau_{{\rm cap}, p}$) is the characteristic time of electron (hole) capture into the well, $G_m$ is the contribution of the $m$-th well to the modal gain, $\alpha_{\rm tot}$ is the total loss, $R_{m}^{(2D)}$ and $R_{{\rm sp}\,m}^{(2D)}$ are the total interband and spontaneous recombination rates, and $\beta$ is the fraction of spontaneous emission going into the lasing mode.

The total loss $\alpha_{\rm tot}$ is the sum of external and internal contributions $\alpha_{\rm ext}$, $\alpha_{\rm int}$, where $\alpha_{\rm int}$ can be further divided into the lattice absorption coefficient $\alpha_{\rm lat}$ and the free-carrier absorption coefficient $\alpha_{\rm FCA}$. The latter was assumed to be proportional to the carrier densities~\cite{1983/Mroczkowski/JAP/Optical,1996/Li/JAP/Free}:
\begin{eq}{free-carrier_absorption}
\alpha_{\rm FCA} = \sigma_n n + \sigma_p p,
\end{eq}
where $\sigma_n$ and $\sigma_p$ are the corresponding absorption cross sections.

Absorption cross sections calculated for CdTe within the Drude model at 30 $\mu$m and 70 K are $\sigma_n = 2.2 \times 10^{-15}$ cm$^2$ and $\sigma_p = 6.9 \times 10^{-15}$ cm$^2$, in accordance with Refs.~\citenum{1983/Mroczkowski/JAP/Optical,2020/Ushakov/OE/HgCdTe}. They set an upper limit to the CdTe waveguide doping at 
$10^{15} - 10^{16}$ cm$^{-3}$~\cite{2020/Ushakov/OE/HgCdTe}.

Spontaneous recombination rates and modal gain were calculated by considering direct optical transitions. Nonlinear effects were taken into account by the nonlinear gain parameter $\varepsilon_S$:
\begin{eq}{nonlinear_gain}
G\left(n_w^{(2D)},p_w^{(2D)},S^{(3D)}\right) = \frac{G\left(n_w^{(2D)},p_w^{(2D)},0\right)}{1 + \varepsilon_S S^{(3D)}}.
\end{eq}
\section{\label{sec:resonator-heating}Distributed model of the resonator and active region heating}

To find the distribution of photon density across the resonator, we used the Bouguer--Lambert--Beer law for the photon densities $A$, $B$ in two counterpropagating waves taking into account reflections at the facets~\cite{2018/Afonenko/QE/Power}:
\begin{eq}{Bouger-Lambert-Beer}
S^{(2D)} &= A + B,\\
\frac{dA}{dx} &= +(G-\alpha_{\rm int})A,\\ \frac{dB}{dx} &= -(G-\alpha_{\rm int})B,\\
\end{eq}
where the $x$-axis points in the direction of propagation of the first wave.

Modal gain and free-carrier absorption coefficient depend on the carrier densities, which can be found from the drift-diffusion model described in Appendix~\ref{sec:drift-diffusion}. Equations (\ref{eq:Bouger-Lambert-Beer}) require dependencies $G$ and $\alpha_{int}$ on the photon density $S^{(2D)}$.
Instead of solving drift-diffusion equations (\ref{eq:Poisson-continuity}) with given photon density $S^{(2D)}$ we artificially change the lattice absorption coefficient $\alpha_{\rm lat}$ to obtain the given photon density $S^{(2D)}$ and the corresponding distribution of carrier densities. This allows us to obtain the functional dependences $G(S^{(2D)},T, j)$, $\alpha_{\rm int}(S^{(2D)},T,j)$ by calculating $G(\alpha_{\rm lat},T,j)$, $\alpha_{\rm int}(\alpha_{\rm lat},T,j)$, and $S^{(2D)}(\alpha_{\rm lat},T,j)$ within the drift-diffusion model. Actually, the functions $G(S^{(2D)},T,j)$, $\alpha_{\rm int}(S^{(2D)},T,j)$ were approximated by a bi-quadratic interpolation between 9 values calculated at a given current density $j$ and different $T$, $\alpha_{\rm lat}$. 

The evolution of local temperature $T(t,x)$ inside the active region after the start of a pulse was found by using Green's functions of the heat equation. Joule heating in the waveguide (including the active region) and in the cladding layers/substrate were considered separately, with corresponding Green's functions ${\cal G}_{\rm in}(\tau)$, ${\cal G}_{\rm out}(\tau)$~\cite{2018/Afonenko/QE/Power}:
\begin{eq}{temperature}
T(t,x) - T_0 &= \int_0^t {\cal G}_{\rm in}(\tau) w_{\rm heat}(t - \tau, x) d\tau\\
&+ \int_0^t {\cal G}_{\rm out}(\tau) j^2(t - \tau, x) d\tau
\end{eq}

The heat power produced in the waveguide per unit area $w_{\rm heat}$ is given by Joule's law minus the power transferred to the laser radiation:
\begin{eq}{heat_power}
w_{\rm heat}(t, x) &= j(t, x)U(t, x)\\
&- v_g \left[G(t, x) - \alpha_{\rm int}(t, x) \right] S^{(2D)}(t, x) \hbar \omega(t),
\end{eq}
where $j(t, x)$ is the current through the waveguide, $U(t, x)$ is the associated voltage drop, $G(t, x)$ is the modal gain, $\alpha_{\rm int}(t, x)$ is the internal loss, $S^{(2D)}(t, x)$ is the 2D photon density in the lasing mode, and $\hbar \omega(t)$ is the photon energy. The lasing frequency $\omega$ is time-dependent because heating affects the bandgap.

Green's functions ${\cal G}_{\rm in}(\tau)$, ${\cal G}_{\rm out}(\tau)$ were found by numerical solution of the one-dimensional heat equation along the $z$-axis (perpendicular to the heterostructure layers) taking into account all epitaxial layers, the substrate, the solder, and the heat sink. Heat transfer along the $x$-axis (parallel to the layers) was neglected. This approach is justified because the thermal diffusion length during the drive current pulse is much larger than the waveguide thickness, but much smaller than the substrate thickness, heat sink thickness, and resonator length. Assuming the pulse duration is $\tau = 1$ $\mu$s, the thermal diffusion length is $\sim 2 \sqrt{\chi \tau} = 13$ $\mu$m ($\chi$ is the heat transfer coefficient~\cite{1972/Slack/PhysRevB/Thermal}), while the active and emitter region thickness is $\sim 0.2$ $\mu$m (Fig.~\ref{fig:bands_and_concentrations}).
\section{\label{sec:Auger}Calculation of Auger coefficients}
\begin{fig}{Auger}{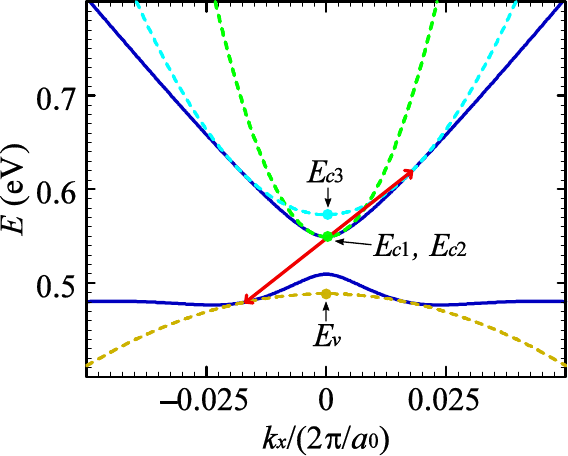}
Threshold CCCH Auger process (red arrows) and parabolic approximation (dashed lines) to the exact bandstructure (solid blue lines). $E_{c1}$, $E_{c2}$, $E_{c3}$, $E_v$ are the energies of the parabolas' extrema.
\end{fig}
In the interband recombination rate $R$ we included both radiative and Auger recombination. In narrow-gap CdHgTe quantum wells the Auger recombination rate is dominated by the CCCH process~\cite{2020/Alymov/ACSPhot/Fundamental}. The strong nonparabolicity of the band dispersion complicates analytical treatment of Auger recombination, so we used parabolic interpolation near the states giving the dominant contribution to the Auger recombination rate (Fig.~\ref{fig:Auger}). In this approximation, two-dimensional Auger coefficient in a deep quantum well reads
\begin{eq}{Auger_coefficient}
C_{nnp}^{(2D)} &= K_{2D/3D}^2 F_{cc\, cv}^2 \left(\frac{2\pi e^2}{\epsilon \epsilon_0 q_{\rm th}}\right)^2 \frac{\pi}{2 \hbar kT}\\
&\times \frac{m_{c1} m_{c2} m_{c3} m_v \left(m_{c1} + m_{c2} - m_{c3} + m_v \right)}{\left(m_c^*\right)^2 m_v^* \left(m_{c1} + m_{c2} + m_v \right)^2}\\ 
&\times \exp\left[-\frac{m_{c3}\left(E_{c1} + E_{c2} - E_{c3} - E_v \right)}{\left(m_{c1} + m_{c2} - m_{c3} + m_v \right) kT}\right].
\end{eq}
Here, $m_c^*$ and $m_v^*$ are the density-of-states effective masses of electrons and holes, $m_{c1}$, $m_{c2}$, $m_{c3}$, $m_v$ are the effective masses of the parabolic bands approximating the exact band dispersion, $E_{c1}$, $E_{c2}$, $E_{c3}$, $E_v$ are the edges of the parabolic bands, $\hbar$ is the reduced Planck's constant, and $F_{cc\, cv}$ is the overlap integral. 

Finite extension of carrier wavefunctions in $z$ direction is taken into account by the factor $K_{2D/3D}$, which can be calculated for sinusoidal envelope wave functions as
\begin{eq}{quasi2D_factor}
K_{2D/3D} = \left( 1+ \frac{q_{\rm th}d}{3} \right)^{-1},
\end{eq} 
where $d$ is the quantum well width, and
\begin{eq}{threshold_momentum}
q_{\rm th} = \frac{1}{\hbar}\sqrt{\frac{2 m_{c3} \left(m_{c2} + m_v \right)^2 \left(E_{c1} + E_{c2} - E_{c3} - E_v \right)}{\left(m_{c1} + m_{c2} + m_v \right)\left(m_{c1} + m_{c2} - m_{c3} + m_v \right)}}
\end{eq}
is the momentum transfer during the threshold Auger process (i. e., the process giving the dominant contribution to the Auger recombination rate).
The calculated matrix element of the Bloch wave function overlap was 0.3, which agrees with the results of \cite{1993/Lopes/SemScienTechn}. Factor $K_{2D/3D}$ was 0.4, which is less than 0.6 calculated by formula (\ref{eq:quasi2D_factor}), since the envelope wave function for electrons  differs from the sinusoidal shape.
For 5.2 nm-thick HgTe/Cd$_{0.6}$Hg$_{0.4}$Te quantum wells used in our simulations, the calculated Auger coefficient $C_{nnp}^{(2D)}$ is $8.3 \times 10^{-12}$ cm$^4$/s at 70 K and $8.7 \times 10^{-12}$ cm$^4$/s at 90 K. The corresponding Auger coefficient defined in terms of 3D carrier densities is $C_{nnp}^{(3D)} = 2 C_{nnp}^{(2D)} d^2/5 = 9.0 \times 10^{-25}$ cm$^6$/s at 70 K and $9.4\times 10^{-25}$ cm$^6$/s at 90 K. (The factor of $2/5$ appears when $n^2 p$ is averaged over the QW width assuming carrier wavefunctions are $\psi(z) = \sqrt{2/d}\sin\left(\pi z/d \right)$.)
\clearpage

\end{document}